\documentstyle[preprint,pre,aps,oldlfont]{revtex}
\begin{document}
\draft
\title{Multiple Transitions to Chaos
       in a Damped Parametrically Forced Pendulum
}
\author{Sang-Yoon Kim
\footnote{Electronic address: sykim@cc.kangwon.ac.kr}
and Kijin Lee
       }
\address{
Department of Physics\\ Kangwon National University\\
Chunchon, Kangwon-Do 200-701, Korea
}
\maketitle

\begin{abstract}
We  study  bifurcations  associated   with  stability  of  the  lowest
stationary
point (SP) of a damped parametrically forced pendulum by varying
$\omega_0$ (the  natural   frequency of  the pendulum)   and  $A$ (the
amplitude
of
the  external  driving  force).  As  $A$  is increased,  the  SP  will
restabilize
after its  instability, destabilize again,  and so  {\it ad infinitum}
for any
given $\omega_0$. Its destabilizations (restabilizations) occur via
alternating       supercritical       (subcritical)    period-doubling
bifurcations
(PDB's)
and pitchfork bifurcations, except  the first destabilization at which
a
supercritical or subcritical bifurcation  takes place depending on the
value
of $\omega_0$. For each case of the supercritical destabilizations, an
infinite sequence of  PDB's follows and leads  to chaos. Consequently,
an
infinite  series  of   period-doubling  transitions  to chaos  appears
with
increasing
$A$.  The  critical  behaviors  at  the  transition  points  are  also
discussed.
\end{abstract}


%
%

\narrowtext

\section{Introduction}
\label{sec:Int}

A   damped    parametrically   forced    pendulum   (DPFP)   with    a
vertically
oscillating
support   is    investigated.   It     can  be     described   by    a
second-order
non-autonomous
ordinary differential equation (ODE) \cite{Landau,Arnold1,Arnold2},
\begin{equation}
{\ddot x} + 2 \pi \gamma {\dot  x} + 2 \pi (\omega_0^2  - A \cos 2 \pi
t)
\sin 2 \pi x = 0,
\label{eq:PFP}
\end{equation}
where $x$ is the angular position, $\gamma$ the damping coefficient,
$\omega_0$ the  undamped natural  frequency of the  unforced pendulum,
and $A$
the amplitude of the external driving force of period one. The overdot
denotes the  differentiation with respect  to time,  and all variables
and
parameters are expressed in dimensionless forms.

The   DPFP,  albeit   looking   simple,   shows  a  richness   in  its
dynamical
behavior.
As the amplitude  $A$ is increased up  to moderate values, transitions
from
periodic  attractors  to  chaotic  attractors  and {\it  vice  versa},
coexistence
of different attractors, transient chaos, and
so on have been found numerically \cite{McLaughlin,Leven1,Arneodo} and
analytically \cite{Leven2}. They have been also observed in real
experiments \cite{Leven3,Leven4}.
However, as $A$ increases further, the DPFP exhibits
new interesting dynamical behaviors not found in previous works,
as will be seen below.

Here we  are interested in  bifurcations associated  with stability of
the
lowest stationary point (SP) with $x=0$ and ${\dot x} =0$ of the DPFP.
The  linear  stability of  the  SP  is  determined  by the  linearized
equation
\begin{equation}
{\ddot x} +  2 \pi \gamma {\dot x}  + 4 \pi^2 (\omega_0^2   - A \cos 2
\pi t)
x = 0,
\label{eq:DMEQ}
\end{equation}
which is a damped Mathieu equation.
For the  undamped case  with $\gamma=0$,  stability properties  of the
Mathieu
equation are given in \cite{Morse,Mathews}.
There exist an infinite number of disconnected
instability regions of the SP in the $\omega_0-A$ plane.
These instability regions may be called
``tongues'',  because  their lower  parts  are  of  tongue shape  (see
Fig.~7-5
in Ref.~\cite{Mathews}).
They can be also labelled by an integer $n$, half of
which correspond, in absence of forcing ($A=0$), to
parametric resonant values  of $\omega_0$, i.e. $\omega_0  = { n \over
2}$.
However, even  a small amount  of damping leads  to the  presence of a
non-zero
threshold value $A_t(n)$ of the amplitude necessary for the occurrence
of
the $n$th-order parametric resonnance \cite{Landau,Arnold1,Arnold2}.
Moreover, $A_t(n)$ grows rapidly with increasing $n$ (see Fig.~100 in
Ref.~\cite{Arnold1}).

We first introduce the Poincar{\'{e}} map for the DPFP in
Sec.~\ref{sec:PM}, and then discuss various bifurcations
associated with stability of periodic orbits.
With  increasing $A$  to  sufficiently large  values,  the bifurcation
behaviors
associated with stability of the SP are investigated
in Sec.~\ref{sec:MT} for a moderately damped case with $\gamma=0.1$.
The   damped    Mathieu Eq.~(\ref{eq:DMEQ})    has   an    infinity of
alternating
stable
and  unstable   $A$  ranges for   any  given  $\omega_0$.   Hence,  as
$A$  is
increased,
the  SP  undergoes  a  cascade  of  ``resurrections''  for  any  given
$\omega_0$,
i.e. it will   restabilize after it loses   its stability, destabilize
again,
and
so {\it ad infinitum}.
Its   restabilizations   occur    through   alternationg   subcritical
period-doubling
bifurcations (PDB's) and pitchfork  bifurcations (PFB's). On the other
hand,
the  destabilizations occur  through  alternating  supercritical PDB's
and
PFB's,
except  the   first  destabilization  at  which   a  supercritical  or
subcritical
bifurcation takes place depending on the value of $\omega_0$.
For  each  case of  the  supercritical  destabilizations, an  infinite
sequence
of PDB's leading to chaos follows. Consequently, an infinite series of
period-doubling  transitions to  chaos  appears  with increasing  $A$,
which was
not found in previous works.
This is in contradistinction to the cases of the one-dimensional (1D)
maps and other damped forced oscillators, for which only one single
period-doubling transition to chaos occurs.
In Sec.~\ref{sec:CB}, we study the critical scaling behaviors at the
transition points. It is found that they are the same as those for the
1D
maps. Finally, a summary is given in Sec.~\ref{sec:Sum}.

\section{Stability of Periodic Orbits, Bifurcations and Lyapunov
         Exponents in the Poincar\'{e} Map}
\label{sec:PM}

In this section, we first discuss stability of period orbits in the
Poincar\'{e} map of the DPFP, using the Floquet theory.
Bifurcations associated with the  stability and Lyapunov exponents are
then
discussed.

The  second-order ODE  (\ref{eq:PFP})  is reduced  to  two first-order
ODE's:
\begin{mathletters}
\begin{eqnarray}
{\dot x} &=& y,  \\
{\dot y} &=& -2 \pi \gamma y - 2 \pi (\omega_0^2  - A \cos 2 \pi t)
\sin 2 \pi x.
\end{eqnarray}
\label{eq:PFP2}
\end{mathletters}
The Poincar\'{e} maps  of an initial point  $z_0 [\equiv (x(0),y(0))]$
can be
computed  by sampling  the points  $z_m$ at  the discrete  time $t=m$,
where
$m=1,2,3,\dots$~. We call the transformation $z_m \rightarrow z_{m+1}$
the
Poincar\'{e} (time-1) map, and write $z_{m+1} = P (z_m)$.

The Poincar\'{e} map $P$ has the inversion symmetry such that
\begin{equation}
SPS(z) = P \;{\rm for\;all\;}z,
\label{eq:Sym}
\end{equation}
where $z=(x,y)$, $S$ is the inversion of $z$, i.e., $S(z)=-z$.
If an  orbit $\{ z_m  \}$ of  $P$ is invariant  under $S$, then  it is
called a
symmetric orbit. Otherwise, it is  called an asymmetric orbit, and has
its
``conjugate'' orbit $S \{ z_m \}$.

We now  study the stability of  a periodic orbit with  period $q$ such
that
$P^q (z_0) =z_0$ but $P^j (z_0) \neq z_0$ for $1 \leq j \leq k-1$.
Here $P^k$ means the $k$-times iterated map.
The linear stability of the $q$-periodic orbit is determined from the
linearized-map matrix $DP^q(z_0)$ of $P^q$ at an orbit point $z_0$.
Using  the   Floquet theory   \cite{Lefschetz1},  the   matrix  $DP^q$
can  be
obtained
by    integrating   the    linearized   differential   equations   for
small
perturbations
as follows.

Let  $z^*(t)=z^*(t+q)$  be  a  solution  lying  on  the  closed  orbit
corresponding
to the $q$-periodic orbit. In order to determine the stability of the
closed orbit, we consider an infinitesimal perturbation
$(\delta x(t), \delta y(t))$ to the closed orbit.
Linearizing Eq.~(\ref{eq:PFP2}) about the closed orbit, we obtain
\begin{equation}
 \left( \begin{array}{c}
         \delta {\dot x}  \\
         \delta {\dot y}
      \end{array}
      \right)
      = J(t)
      \left( \begin{array}{c}
         \delta x  \\
         \delta y
      \end{array}
      \right),
\label{eq:LEQ}
\end{equation}
where
\begin{equation}
 J(t) =
 \left( \begin{array} {cc}
        0 & 1 \\
        -4 \pi^2 (\omega_0^2 - A \cos 2 \pi t) \cos 2 \pi x^*(t) &
        -2 \pi \gamma
        \end{array}
 \right).
\end{equation}
Note that $J$ is a $2 \times 2$ $q$-periodic matrix.
Let $W(t)=(w^1(t),w^2(t))$ be a fundamental solution matrix  with
$W(0) = I$. Here $w^1(t)$ and $w^2(t)$ are two
independent solutions expressed in column vector forms, and $I$ is the
$2 \times 2$ unit matrix. Then a general solution of the $q$-periodic
system has the following form
\begin{equation}
 \left( \begin{array}{c}
         \delta x(t)  \\
         \delta y(t)
      \end{array}
      \right)
      = W(t)
      \left( \begin{array}{c}
         \delta x(0)  \\
         \delta y(0)
      \end{array}
      \right).
\label{eq:FSM}
\end{equation}
Substitution of Eq.~(\ref{eq:FSM}) into Eq.~(\ref{eq:LEQ}) leads to
an initial-value problem to determine $W(t)$
\begin{equation}
{\dot W(t)} = J(t) W(t), \;\;W(0)=I.
\label{eq:FSMEQ}
\end{equation}
It  is  clear   from  Eq.~(\ref{eq:FSM})  that  $W(q)$   is  just  the
linearized-map
matrix   $DP^q(z_0)$.   Hence   the  matrix   $DP^q$    is  calculated
through
integration
of Eq.~(\ref{eq:FSMEQ}) over the period $q$.

The characteristic equation of the linearized-map matrix
$M (\equiv DP^q)$ is
\begin{equation}
\lambda^2 - {\rm tr}M \, \lambda + {\rm det} \, M = 0,
\end{equation}
where ${\rm tr}M$ and ${\rm det}M$ denote the trace and determinant of
$M$,
respectively.   The eigenvalues,   $\lambda_1$   and $\lambda_2$,   of
$M$ are
called
the Floquet stability multipliers. As shown in \cite{Lefschetz2},
${\rm det}\,M$ is calculated from a formula
\begin{equation}
{\rm det}\,M = e^{\int_0^q {\rm tr}\,J dt}.
\label{eq:Det}
\end{equation}
Substituting the trace  of $M$ (i.e., ${\rm  tr} J = -  2 \pi \gamma$)
into
Eq.~(\ref{eq:Det}), we obtain
\begin{equation}
{\rm det}\,M = e^{-2 \pi \gamma q}.
\end{equation}
Hence, the Poincar{\' e} map $P$ is a two-dimensional (2D) dissipative
map
with a constant Jacobian determinant  (less than unity), like the H{\'
e}non
map \cite{Henon}.

The pair of  stability multipliers of a periodic  orbit lies either on
the
circle  of radius  $e^{-\pi \gamma  q}$, or  on the  real axis  in the
complex
plane. The  periodic orbit  is stable  only when both  multipliers lie
inside
the unit circle. We first  note that they never cross the unit circle,
and
hence Hopf bifurcations do not occur.
Consequently, it can lose its stability only when a multiplier
decreases (increases) through $-1$ $(1)$ on the real axis.

A more  convenient real quantity  $R$, called the  residue and defined
by,
\begin{equation}
R \equiv { {1 + {\rm det}M - {\rm tr}M} \over {2(1+{\rm det}M)}},
\end{equation}
was  introduced in  \cite{Kim} to  characterize stability  of periodic
orbits
in 2D dissipative maps with constant Jacobian determinants.
A periodic  orbit  is stable  when $  0  < R <1   $; at both  ends  of
$R=0$ and
$1$,
the stability multipliers $\lambda$'s are $1$ and $-1$, respectively.
When  $R$ decreases  through  $0$ (i.e.,  $\lambda$  increases through
$1$), the
periodic orbit loses its stability via saddle-node or pitchfork or
transcritical bifurcation.   On the other   hand, when   $R$ increases
through
$1$
(i.e., $\lambda$ decreases through $-1$), it becomes unstable via
PDB, also referred to as a flip or subharmonic
bifurcation.  For each  case  of the  PFB's  and PDB's,  two  types of
supercritical
and subcritical
bifurcations occur. For more details on bifurcations, refer to
Ref.~\cite{Gukenheimer}.

Lyapunov exponents of  an orbit $\{ z_m \}$  in the Poincar{\'{e}} map
$P$
characterize the mean exponential rate of divergence of nearby orbits
\cite{Lichtenberg}.
There exist two Lyapunov exponents $\sigma_1$ and $\sigma_2$
($\sigma_1 \geq  \sigma_2$) such  that $\sigma_1  + \sigma_2 =  -2 \pi
\gamma$,
because the linearized Poincar{\'{e}} map $DP$ has a constant Jacobian
determinant, det$DP = e^{-2 \pi \gamma}$.
We choose  an initial perturbation  $\delta z_0$ to  the initial orbit
point
$z_0$ and iterate the linearized map $DP$ for
$\delta z$ along the orbit to obtain the magnitute $d_m$
$(\equiv |\delta z_m|)$ of $\delta z_m$.
Then, for  almost all infinitesimally-small  initial perturbations, we
have
the largest Lyapunov exponent $\sigma_1$ given by
\begin{equation}
\sigma_1 = \lim_{m \rightarrow \infty}
   {1 \over m}  \ln {d_m \over d_0}.
\end{equation}
If $\sigma_1$ is positive, then the orbit is called a chaotic orbit;
otherwise, it is a regular orbit.

\section{Multiple Period-Doubling Transitions to Chaos}
\label{sec:MT}

In  this section,  by varying  two parameters  $\omega_0$ and  $A$, we
study
bifurcations associated with stability of the SP of the DPFP for a
moderately  damped  case with  $\gamma=0.1$.  It  is  found that  with
increasing
$A$,  the   SP  undergoes   an  infinite  series   of  period-doubling
transitions
to chaos for any given $\omega_0$.  This is in contrast to 1D maps and
other
damped forced oscillatiors with only single period-doubling transition
to
chaos.

The stability diagram of the SP is given in Fig.~\ref{SD}. There exist
an
infinity of disconnected instability
regions  in  the $\omega_0$-$A$  plane,  which  are  separated by  one
connected
stability   region.    The   instability    regions   may  be   called
``tongues'',
because
their  lower parts   are  tongue-shaped. They  can   be also  labelled
by an
integer
$n$,   half   of   which   correspond   to   the parametric   resonant
values  of
$\omega_0$
(i.e., $\omega_0 = {n \over 2})$ in absence of forcing ($A=0$) for
the                  undamped                 case                  of
$\gamma=0$
\cite{Landau,Arnold1,Arnold2,Morse,Mathews}.
However, even a small amount  of damping results in a non-zero minimal
value
$A_t(n)$  of  the  amplitude  necessary  for  the  occurrence  of  the
$n$th-order
parametric   resonance   \cite{Landau,Arnold1,Arnold2}.   Furthermore,
$A_t(n)$
grows rapidly with increasing $n$ (see Fig.~\ref{SD}).
Hereafter, each tongue of order $n$ is denoted by $T_n$.

With increasing  $A$, each tongue  $T_n$ is twisted  to the  left, and
lies
above
the tongue $T_{n-1}$. In such  a way, tongues pile up successively, as
shown
in   Fig.~\ref{SD}.  Consequently,   there   exist  an   infinity   of
alternating
stable and
unstable  $A$  ranges for  any  given  $\omega_0$.  Hence,  as $A$  is
increased,
the SP will restabilize after it loses its instability, destabilize
again, and so {\it ad infinitum} for any given $\omega_0$.
Such ``resurrection'' mechanisms are given below.

Bifurcation  behaviors at  the tongue  boundaries are  investigated in
details.
They depend on whether the tongue-order $n$ is odd or even.
At the tongue  boundaries of odd (even) order $n$,  the residue of the
SP is
$1$  $(0)$.    Consequently, PDB's  and    PFB's  occur  when   tongue
boundaries of
odd
and even order
$n$ are  crossed, respectively. For  example, the  boundaries of $T_1$
and
$T_3$ in Fig.~\ref{SD} are PDB  curves, while the boundary of $T_2$ is
a
PFB curve. For the cases of PDB's and PFB's, there are two types of
supercritical and  subcritical bifurcations, which  occur depending on
where
tongue boundaries are crossed.  A saddle-node bifurcation (SNB) curve,
at
which a  pair of stable and  unstable orbits with period  $2$ $(1)$ is
born,
touches each  tongue boundary  of odd (even)  order $n$ at  a boundary
point
$[\omega_b(n), A_b(n)]$, and decomposes it into the
supercritical   and subcritical   parts.  As   an   example, see   the
three SNB
curves,
denoted  by  dash-dotted curves,  touching  the  boundaries of  $T_1$,
$T_2$, and
$T_3$,  respectively, in  Fig.~\ref{SD}. On  the  lower left  part  of
each
tongue
boundary,  denoted  by  a  solid  curve, a  supercritical  bifurcation
occurs.
The    remaining    subcritical    boundary    curve  starting    from
$[\omega_b(n),
A_b(n)]$
first   goes    to   the   right,    but   it   turns   left    at   a
point
$[\omega_t(n),A_t(n)]$.
It  consists of  two types  of subparts,  denoted by  short-dotted and
dashed
curves, on which a subcritical bifurcation takes place.
On   the  subcritical   segment   with  $\omega_b(n)   <   \omega_0  <
\omega_t(n)$, the
SP absorbes  an unstable  orbit born  at a  dash-dotted SNB  curve and
loses its
stability. On the other hand, the stable orbit born by the same SNB
undergoes  an  infinite   series  of  PDB's  leading   to  chaos.  The
accumulation
points of such PDB's are denoted by open circles in Fig.~\ref{SD}.

When the SP loses its stability via supercritical PDB's and PFB's, the
system is asymptotically attracted to periodic attractors (born by the
supercritical  bifurcations)  with the  doubled  period  and the  same
period,
respectively. However, for the subcritical bifurcation cases, the
asymptotic states just after the instability of the SP may be periodic
or chaotic, depending on which subparts of the subcritical boundaries
are  crossed.  In  Fig.~\ref{AS},  we  fix  different  $\omega_0$  and
increase $A$
to
cross different subparts of a subcritical boundary of $T_1$.
When a short-dotted boundary curve is crossed, the asymptotic state
becomes   periodic [see   Fig.~2(a)],  because   the SP   jumps   to a
periodic
attractor
born by an SNB after its instability.
For this periodic case, with increasing $A$ an infinite sequence of
supercritical PDB's leading to small-scale chaos follows.
However, when a dashed boundary is crossed, large-scale full chaos
appears via intermittency \cite{Intermittency}, and hence the
asymptotic state becomes chaotic [see Fig.~2(b)].

With increasing $A$ to sufficiently large values, the bifurcation
behaviors  associated with  stability of  the  SP are  investigated in
details
for many values of $\omega_0$.
For  a  given  $\omega_0$,  the   restabilizations  of  the  SP  occur
via
alternating
subcritical  PDB's  and  PFB's   with  increasing  $A$,  as  shown  in
Fig.~\ref{SD}.
On the
other   hand,  the   destabilizations  take   place   via  alternating
supercritical
PDB's  and  PFB's,  except   the  first  destabilization  at  which  a
supercritical
or   subcritical bifurcation   occurs  depending   on   the value   of
$\omega_0$
(e.g.,
for $\omega_0=0.5$ $(0.65),$ the first destabilization occurs via
supercritical (subcritical) PDB). For each case of the supercritical
destabilizations, an infinite  sequence of supercritical PDB's leading
to a
pair of chaotic  attractors follows and ends  at a finite accumulation
point.
In each tongue, such accumulation points of PDB's, denoted by  solid
circles in Fig.1, seem to form a smooth critical line .
Consequently, an infinite series of
period-doubling transitions to chaos appears with increasing $A$.
This is in contradistinction to the 1D maps and other damped forced
oscillators, in which only single period-doubling transition to chaos
occurs.

As  an   example  of   the  multiple period-doubling   transitions  to
chaos,
consider
the case $\omega_0=0.5$. A bifurcation diagram along the vertical line
$\omega_0=0.5$  is shown  in Fig.~\ref{BD1}.  Through  a supercritical
PDB, the
SP
loses its stability at its first destabilization point
$A_d(1)=0.100\,218\, \cdots$, and  a symmetric orbit of  period $2$ is
born.
Unlike  the case  of the  SP, the  symmetric 2-periodic  orbit becomes
unstable
by a symmetry-breaking supercritical PFB,  which leads to the birth of
a
conjugate pair of asymmetric orbits with period $2$. (For the sake of
convenience, only one asymmetrical orbit of period $2$ is shown in
Fig.~3 \cite{rem}.) However, as $A$ is
further increased, an infinite sequence of supercritical PDB's follows
and
ends at its accumulation point
$A^*_1$ $(=0.357\,709\,84\,  \cdots)$. The  critical scaling behaviors
of
period doublings near the critical point $A^*_1$ are the same as those
for the 1D maps, as will be seen in Sec.~\ref{sec:CB}.

After the period-doubling transition to chaos, a conjugate pair of
small chaotic atttractors with positive largest Lyapunov exponent
$\sigma_1$ appear. As $A$ is increased, the different parts
of a chaotic attractor coalesce and form larger pieces. For example,
the chaotic attractor with $\sigma_1 \simeq 0.091$ shown in
Fig.~\ref{Lexp}(a) seems to be composed of four distinct pieces for
$A=0.3579$. As shown in Fig.~\ref{Lexp}(b),
these pieces coalesce to form two large pieces with $\sigma_1=0.158$
for $A=0.3582$. However, beyond some critical point
$A_c(1)$ $(  \simeq 0.3586)$, the chaotic  attractor becomes unstable,
and
the system is asymptotically attracted to a rotational orbit of period
1
born by an SNB.
For  $A>A_c(1)$,   the  DPFP  continues  to   exhibit  rich  dynamical
behaviors.
With increasing $A$, birth of new periodic attractors via SNB's,
transitions from periodic attractors to chaotic attractors
and {\it vice versa}, coexistence of different attractors,
and so  on are found until  the SP restabilizes. (For  more details on
such
dynamical behaviors, refer to previous works
\cite{McLaughlin,Leven1,Arneodo,Leven2,Leven3,Leven4}.)
However,  with   increasing  $A$   further,  the  DPFP   exhibits  new
interesting
dynamical behaviors not previously found.

When the dashed subcritical boundary of $T_1$ is crossed at the first
restabilization   point   $A_r(1)$   $(=3.150\,509\,   \cdots)$,   the
 SP
restabilizes
via subcritical PDB. An ``inverse'' process of the case of Fig.~2(b)
occurs. There exists large-scale full chaos below $A_r(1)$. When $A$
increases through  $A_r(1)$, the  large chaotic  attractor disappears,
and
the   restabilization  of   the  SP   occurs  with  birth  of   a  new
unstable
2-periodic
orbit. The residue $R$ of the SP decreases monotonically
from one, and becomes zero at the second destabilization point
$A_d(2)$ $(=3.224\,230\,  \cdots)$ on  the supercritical PFB  curve of
$T_2$.

A  second   bifurcation  diagram   for  $\omega_0=0.5$  is   shown  in
Fig.~\ref{BD2}.
The SP becomes unstable via symmetry-breaking supercritical PFB at its
second destabilization point $A_d(2)$, which results in the birth of a
conjugate pair of asymmetric orbits with period $1$.
With  further  increase   of  $A$,  a  second   infinite  sequence  of
supercritical
PDB's follows and ends at its accumulation point
$A^*_2$ $(=3.263\,703\,15\, \cdots)$. The critical scaling behaviors
of period  doublings near  $A=A^*_2$ are  the same  as those  near the
first
accumulation point $A^*_1$.
After the second period-doubling transition to chaos, a conjugate pair
of
small  chaotic  attractors  also  appears.  They  persist  until  some
critical
point $A_c(2)$ $(\simeq 3.263\,862)$, beyond which the
system  is asymptotically   attracted to  an oscillating  $2$-periodic
orbit
born
via SNB.
As in the tongue of order $1$, the DPFP exhibits diverse dynamical
behaviors  such  as  transitions  between  the  periodic  and  chaotic
attractors
and  the coexistence  of different  attractors  in the  region between
$A_c(2)$
and  the   second  restabilization   point  $A_r(2)$  $(=10.093\,985\,
\cdots)$.

When the dashed subcritical boundary of $T_2$ is crossed at $A_r(2)$,
a subcritical PFB occurs. Consequently, the SP restabilizes with birth
of a  new unstable orbit of  period $1$. As $A$  is further increased,
the
residue $R$ of the SP monotonically increases, and becomes one at the
third destabilization point $A_d(3)$ $(=10.097\,583\, \cdots)$ on the
supercritical PDB curve of $T_3$. Since the order of $T_3$ is odd, the
subsequent bifurcation   behaviors in   $T_3$ are  the   same as those
for the
case
of  $T_1$.  That  is,   a  third infinite   sequence  of supercritical
PDB's,
leading
to  a pair  of  small  chaotic attractors,  follows  and  ends at  its
accumulation
point $A^*_3$ $(=10.099\,660\,93\, \cdots)$.
This   third  bifurcation    diagram   for   $\omega_0  =     0.5$  is
given  in
Fig.~\ref{BD3}.
The critical  scaling behaviors of  period doublings near  $A^*_3$ are
also
the
same as those near $A^*_1$, as will be seen in the next section.

We have also  studied many other cases  with different $\omega_0$, and
found
multiple period-doubling transitions to chaos with increasing $A$.
Such   accumulation   points  are   denoted   by   solid  circles   in
Fig.~\ref{SD}.
In each tongue of order $n$, they form a smooth critical line
$A^*_n(\omega_0)$. Since the range of $\omega_0$ is
$0<  \omega_0 <  \omega_b(n)$, each  critical line  of order  $n$ ends
inside
the tongue  with  order  $n$. As mentioned   above, a stable  periodic
orbit,
born
at a dash-dotted SNB curve, also undergoes an infinite sequence of
supercritical PDB's. The accumulation points of such PDB's, denoted by
open  circles  in  Fig.~\ref{SD},  form  another  critical  line.  The
two
different
critical   lines     joins   at    a   point     with   $\omega_0    =
\omega_b(n)$.
Consequently,
each critical line  of order $n$ extends to the  outside of the tongue
of
order $n$.

\section{Critical Behaviors of Period-doubling Bifurcations}
\label{sec:CB}

In this section,  we study the critical behaviors  (CB's) of PDB's for
many
values  of $\omega_0$.  The  orbital scaling  behavior  and  the power
spectra of
the  periodic  orbits  born   via  PDB's as   well  as   the parameter
scaling
behavior
are particularly investigated. The CB's for all cases
studied are found to be the same as those for the 1D maps.

As an example, we consider the case $\omega_0=0.5$. The first three
period-doubling transition points $A^*_i$'s $(i=1,2,3)$ are shown in
Fig.~\ref{SD}. Only the CB's near $A^*_1$ are given below, because the
CB's
at the three transition points are the same. For this case, we follow
the periodic  orbits of period $2^k$  up to level  $k=8$. As explained
above,
for $A=A_d(1)$,  the SP becomes  unstable via supercritical  PDB and a
new
symmetric  2-periodic   orbit  appears  (see   Fig.~3).  However,  the
symmetric
orbit    of     period    $2$     loses    its   stability     by    a
supercritical
symmetry-breaking
PFB   at  $A=0.335\,257\,   \cdots$~.   As  a   result,   a  conjugate
pair  of
asymmetric
$2$-periodic  orbits  appears.  As  $A$  is  further  increased,  each
asymmetrical
orbit   with  period   $2$    undergoes an    infinite   sequence   of
supercritical
PDB's,
ending  at  its  accumulation  point  $A^*_1$. Table  \ref{PSB}  gives
the
$A$-values
at which  the supercritical  PDB's take  place; at $A_k$,  the residue
$R_k$ of
an asymmetric orbit of period $2^k$ is one.
The sequence  of $A_k$  converges asymptotically geometrically  to its
limit
value $A^*_1$ with ratio $\delta$:
\begin{equation}
\delta_k = {{A_k - A_{k-1}} \over {A_{k+1} - A_k}} \rightarrow \delta.
\end{equation}
The sequence  of $\delta_k$  is also  listed in Table  \ref{PSB}. Note
that its
limit   value  $\delta$   $(\simeq   4.67)$  agrees  well  with   that
$(=4.669
\cdots)$
for a 1D map $x_{m+1} = f(x_m )$ with a single quadratic maximum $x^*$
\cite{Feigenbaum1}.
We also obtain the value of $A^*_1$ $(=0.357\,709\,845\,3)$
by superconverging the sequence of $\{ A_k \}$ \cite{MacKay}.

For the  1D map $f$,  consider a $2^k$-periodic  orbit point $x^{(k)}$
nearest
to
the maximum  point $x^*$  when the  orbit becomes unstable.  Then, the
sequence
of   $x^{(k)}$ also  converges  asymptotically  geometrically  to  the
maximum
point
$x^*$ with  ratio $\alpha=-2.502\cdots$  \cite{Feigenbaum1}. Note that
the
region near   the maximum point $x^*$   is the most   rarified region,
because
the
distance     between    $x^{(k)}$     and     its     nearest    orbit
point
$f^{2^{k-1}}(x^{(k)})$
is  maximum. Hence,  for the   case of  the Poincar{\'e}  map $P$,  we
first
locate
the most rarified region by choosing an orbit point $z^{(k)}$
$[=(x^{(k)},y^{(k)})]$ which has
the   largest     distance   from      its   nearest   orbit     point
$P^{2^{k-1}}(z^{(k)})$
for
$A=A_k$. The two   sequences $\{ x^{(k)} \}$ and  $\{  y^{(k)} \}$ are
listed
in
Table \ref{OS}.
Note that  they converge  asymptotically geometrically to  their limit
values
$x^*$ and $y^*$ with the 1D ratio $\alpha$.
\begin{equation}
\alpha_{x,k} = { {x^{(k)} - x^{(k-1)}} \over {x^{(k+1)} - x^{(k)}} }
\rightarrow \alpha, \;\;
\alpha_{y,k} = {{y^{(k)} - y^{(k-1)}} \over
{y^{(k+1)} - y^{(k)}}} \rightarrow \alpha.
\end{equation}
The values of $x^*$ $(=0.091\,126)$ and $y^*$ $(=0.735\,292)$ are also
obtained by superconverging the sequences of $x^{(k)}$ and $y^{(k)}$,
respectively.

We also study the power spectra of the $2^k$-periodic orbits
$(k=1,\ldots,8)$ at the PDB points $A_k$.
Consider the orbit of level $k$ whose period is $q=2^k$,
$\{ z^{(k)}_m=(x^{(k)}_m,y^{(k)}_m), \; m=0,1,\ldots,q-1 \}$. Then,
the $j$th Fourier component of this $2^k$-periodic orbit is given by
\begin{equation}
z^{(k)}(\omega_j) =   {1  \over q}  \sum_{m=0}^{q-1}   z^{(k)}_m e^{-i
\omega_j
m},
\end{equation}
where $\omega_j = 2 \pi j / q$, and $j=0,1,\ldots,q-1$.
The power spectrum $P^{(k)}(\omega_j)$ of level $k$ defined by
\begin{equation}
P^{(k)}(\omega_j) = |z^{(k)}(\omega_j)|^2,
\end{equation}
has discrete peaks at $\omega = \omega_j$.
In the  power spectrum  of the  next $(k+1)$ level,  new peaks  of the
$(k+1)$th
generation appear at odd harmonics of the fundamental frequency,
$\omega_j = 2 \pi (2j+1) / 2^{(k+1)}$ $(j=0, \ldots, 2^k -1)$.
To  classify  the  contributions  of  successive PDB's  in  the  power
spectrum of
level $k$, we write
\begin{equation}
P^{(k)}=       P_{00}       \delta(\omega)      +       \sum_{l=1}^{k}
\sum_{j=0}^{2^{(l-1)}-1}
P^{(k)}_{lj} \delta(\omega- \omega_{lj}),
\end{equation}
where  $P^{(k)}_{lj}$  is   the  height  of  the $j$th   peak  of  the
$l$th
generation
appearing at $\omega=\omega_{lj}$ $(\equiv 2 \pi (2j+1) / 2^l)$.
As an example, see the   power spectrum $P^{(8)}(\omega)$ of level $8$
shown
in
Fig.~\ref{PS}.  The  average   height  of  the  peaks  of   the  $l$th
generation is
given by
\begin{equation}
\phi^{(k)}(l)   =    {1   \over    2^{(l-1)}}   \sum_{j=0}^{2^{l-1}-1}
P_{lj}^{(k)}.
\end{equation}
It is of interest whether the sequence of the ratios of the successive
average heights
\begin{equation}
 2 \beta^{(k)}(l) = \phi^{(k)}(l) / \phi^{(k)}(l+1),
\end{equation}
converges. The  ratios are  listed  in  Table \ref{PSS}. They  seem to
approach
a
limit value,   $2 \beta   \simeq 21$,   which agrees  well  with  that
$(=20.96
\cdots)$
for the 1D map \cite{Rudnick}.

\section{Summary}
\label{sec:Sum}
Bifurcations  associated  with  stability  of   the  SP  of  the  DPFP
are
investigated
by varying two parameters $\omega_0$ and $A$.
As $A$ is increased, the SP undergoes an infinite sequence of
alternating  restabilizations  and   destabilizations  for  any  given
$\omega_0$.
The   restabilization   and   destabilization    mechanisms   are also
given  in
details.
A  new  finding   is  that  an  infinite   series  of  period-doubling
transitions to
chaos appears  with increasing  $A$, which  was not found  in previous
works.
This is in contradistinction to the cases of the 1D maps and other
damped forced oscillators with only single period-doubling transition.
The critical  scalings  at the  transition points are   also found  to
be the
same
as those of the 1D maps.

\acknowledgments
This  work  was supported  by  the  Basic  Science Research  Institute
Program,
Ministry of Education, Korea, Project No. BSRI-95-2401.

%
%

\begin{table}
\caption{ Asymptotically geometric convergence of the parameter values
for
          successive supercritical PDB's from an asymmetric 2-periodic
          orbit.
        }
\label{PSB}
\begin{tabular}{ccc}
$k$ & $A_k$ & $\delta_k$  \\
\tableline
1 & 0.354\,163\,288\,011 &        \\
2 & 0.357\,022\,317\,174 &   5.286 \\
3 & 0.357\,563\,141\,135 &   4.692 \\
4 & 0.357\,678\,400\,212 &   4.665 \\
5 & 0.357\,703\,107\,281&   4.666\\
6 & 0.357\,708\,401\,983&   4.668 \\
7 & 0.357\,709\,536\,272&   4.670 \\
8 & 0.357\,709\,779\,136&
\end{tabular}
\end{table}

\begin{table}
\caption{   Asymptotically  geometric   convergence  of   the  orbital
sequences
          $\{ x_k \}$ and $\{ y_k \}$.
        }
\label{OS}
\begin{tabular}{cccccc}
$k$ & $x_k$ & $\alpha_{x,k}$ & $y_k$ & $\alpha_{y,k}$   \\
\tableline
1 & 0.094\,410\,516 &          &  0.719\,956\,679 &         \\
2 & 0.088\,901\,931 &   -1.933 &  0.738\,357\,722 &  -3.935 \\
3 & 0.091\,750\,680 &   -3.085 &  0.733\,681\,177 &  -2.156 \\
4 & 0.090\,827\,396 &   -2.261 &  0.735\,850\,056 &  -2.717 \\
5 & 0.091\,235\,660 &   -2.635 &  0.735\,051\,829 &  -2.398 \\
6 & 0.091\,080\,705 &   -2.436 &  0.735\,384\,746 &  -2.558 \\
7 & 0.091\,144\,315 &   -2.538 &  0.735\,254\,611 &  -2.474 \\
8 & 0.091\,119\,256 &          &   0.735\,307\,206 &
\end{tabular}
\end{table}

\begin{table}
\caption{            Sequence             $2           \beta^{(k)}(l)$
$[\equiv
\phi^{(k)}(l)/\phi^{(k)}(l+1)]$
of the ratios of the successive average heights.
        }
\label{PSS}
\begin{tabular}{cccccc}
\multicolumn{1}{c}{$k$} & \multicolumn{5}{c}{$l$} \\
 & $3$ & $4$ & $5$ & $6$ & $7$  \\
\tableline
6 & 19.8 & 22.5 & 21.1 & & \\
7 & 19.8 & 22.1 & 21.2 & 21.5 & \\
8 & 19.8 & 22.0 & 20.7 & 21.6 & 21.4
\end{tabular}
\end{table}

\begin{figure}
\caption{Stability  diagram of  the SP  of  the DPFP.  There  exist an
infinity
of ``tongues'' $T_n$ of instability regions. For each tongue,
a  supercritical  bifurcation  occurs  on  the solid  boundary  curve,
whereas a
subcritical  bifurcation  takes  place  on  the  remaining  dashed  or
short-dotted
boundary  curve.  There  are  also  SNB  curves  touching  the  tongue
boundaries,
which are denoted  by the dash-dotted curves.  The accumulation points
of
PDB's, denoted by solid and open circles, form critical lines.
For other details, see the text.
     }
\label{SD}
\end{figure}

\begin{figure}
\caption{Asymptotic  states  after  the  instability  of  the  SP  via
subcritical
         bifurcations. A  pair of symmetric stable  and unstable orbit
of
         period 2 are born via an SNB. The $x$-positions of the stable
and
         unstable orbits are denoted by the solid and
         dashed curves, respectively. At  a subcritical PDB point, the
SP,
         whose $x$-position is denoted by the dotted line, loses its
         stability by absorbing the unstable $2$-periodic orbit. After
its
         instability, (a) the SP jumps to the stable $2$-perodic orbit
         for $\omega_0=0.55$,   whereas (b)   large-scale  full  chaos
appears
for
         $\omega_0=0.6832$. Note also that for each case, the stable
         $2$-periodic orbit  undergoes an  infinite sequence  of PDB's
leading
         to small-scale chaos.
     }
\label{AS}
\end{figure}

\begin{figure}
\caption{ First bifurcation diagram for $\omega_0=0.5$.
          The  SP2  and  ASP2  denote  the stable  $A$-ranges  of  the
symmetric
          and asymmetric orbits of period 2, respectively. The PN also
          designates the  stable $A$-range of  the asymmetric periodic
orbit
          with period N (N=$4,8,16$).
     }
\label{BD1}
\end{figure}

\begin{figure}
\caption{   Chaotic  attractors   after   the   first  period-doubling
transition
          to chaos. (a) For $A=0.3579$, the chaotic attractor with the
          largest  Lyapunov   exponent  $\sigma_1   \simeq  0.091$  is
composed of
          four pieces. (b) These pieces merge to form two large pieces
          with $\sigma_1 \simeq 0.158$ for $A=0.3582$.
     }
\label{Lexp}
\end{figure}

\begin{figure}
\caption{ Second bifurcation diagram for $\omega_0=0.5$.
          The  ASP1  and  PN  denote  the  stable  $A$-ranges  of  the
asymmetric
          orbit of period 1 and N (N=$2,4,8,16$), respectively.
     }
\label{BD2}
\end{figure}

\begin{figure}
\caption{ Third bifurcation diagram for $\omega_0=0.5$.
          The  SP2  and  ASP2  denote  the stable  $A$-ranges  of  the
symmetric
          and asymmetric orbits of period 2, respectively. The PN also
          designates the  stable $A$-range of  the asymmetric periodic
orbit
          with period N (N=$4,8,16$).
     }
\label{BD3}
\end{figure}

\begin{figure}
\caption{ Power spectrum $P^{(8)}(\omega)$ of level $8$ for
          $A=A_8$ $(=0.357\,709\,779\,136)$
     }
\label{PS}
\end{figure}


\begin{references}
\bibitem{rem} For each asymmetric case, only one of the two conjugate
 asymmetric orbits is shown in  a bifurcation diagram, or in the phase
plane.
\end{references}
\end{document}